\documentclass[12pt]{article}

\RequirePackage[OT1]{fontenc}
\usepackage[authoryear]{natbib}
\RequirePackage[colorlinks,citecolor=blue,urlcolor=blue]{hyperref}
\usepackage[english]{babel}
\usepackage{amsthm}
\usepackage{amsmath}
\usepackage{amsfonts}
\usepackage{amssymb}
\usepackage{graphicx}
\usepackage{enumerate}
\usepackage{color}
\usepackage{array}

\def\calB{{\mathcal B}}
\def\calM{{\mathcal M}}
\def\calN{{\mathcal N}}
\def\calS{{\mathcal S}}
\def\calT{{\mathcal T}}
\def\calX{{\mathcal X}}
\DeclareMathSymbol{\Real}{\mathbin}{AMSb}{"52}
\def\tt#1{{\texttt{#1}}}
\def\bar#1{{\overline{#1}}}
\def\hat#1{{\widehat{#1}}}

\begin{document}
\thispagestyle{empty}
\baselineskip=28pt
\vskip 5mm
\begin{center} {\Large{\bf Editorial: 
EVA 2019 data competition on spatio-temporal prediction of Red Sea surface temperature extremes}}
\end{center}

\baselineskip=12pt
\vskip 5mm

\begin{center}
\large
Huser, Rapha\"el$^1$
\end{center}

\footnotetext[1]{
\baselineskip=10pt Computer, Electrical and Mathematical Sciences and Engineering (CEMSE) Division, King Abdullah University of Science and Technology (KAUST), Thuwal 23955-6900, Saudi Arabia. E-mail: raphael.huser@kaust.edu.sa}

\baselineskip=17pt
\vskip 4mm
\centerline{\today}
\vskip 6mm

\begin{center}
{\large{\bf Abstract}}
\end{center}

Large, non-stationary spatio-temporal data are ubiquitous in modern statistical applications, and the modeling of spatio-temporal extremes is crucial for assessing risks in environmental sciences among others. While the modeling of extremes is challenging in itself, the prediction of rare events at unobserved spatial locations and time points is even more difficult. In this editorial, we describe the data competition that was organized for the 11th international conference on Extreme-Value Analysis (EVA 2019), for which several teams modeled and predicted Red Sea surface temperature extremes over space and time. After introducing the dataset and the goal of the competition, we disclose the final ranking of the teams, and we finally discuss some interesting outcomes and future challenges.
\baselineskip=16pt

\par\vfill\noindent

\pagenumbering{arabic}
\baselineskip=24pt

\newpage

\section{Introduction}
The 11th international conference on Extreme-Value Analysis (EVA 2019), which successfully took place in Zagreb, Croatia, on July 1--5, 2019, gathered experts in mathematical statistics and probability theory to present and discuss recent research advances covering the whole spectrum of extreme-value theory and its application, with topics as varied as, e.g., \emph{Extremes and machine learning}, \emph{Risk analysis in insurance}, \emph{Spatial extremes}, \emph{Detection and attribution of climate change}, \emph{Topological extremes}, \emph{Extremes and graphs}, \emph{Extremes and climate physics}, \emph{Prediction of extremes}, \emph{Time series extremes}, among others. In particular, modeling and accurately predicting the magnitude and the extent of extreme events that take place over space and time is key to assessing risks in a number of applied disciplines, including environmental sciences. Modeling spatio-temporal extremes requires flexible yet parsimonious models with computationally feasible inference. Furthermore, the prediction of unprecedented extreme events that exceed the observed maximum may involve large uncertainties, hence requiring efficient yet resilient estimation approaches. This can become very challenging when the dataset at hand is high-dimensional and non-stationary, as is often the case in modern real data applications.

The literature on spatio-temporal extremes is growing rapidly and several approaches have already been proposed to model extremal dependence and make inference. One possibility, which has found widespread interest because of its solid theoretical foundations, is to model extremes defined as block maxima using max-stable processes \citep[see, e.g.,][]{Padoan.etal:2010,Reich.Shaby:2012a,Opitz:2013a,Huser.Genton:2016,Oesting.etal:2017,Vettori.etal:2019}. However, the models are usually limited to low dimensions; see \citet{Castruccio.etal:2016}, \citet{Huser.etal:2019} and the reviews \citet{Davison.etal:2012,Davison.etal:2019}. Alternatively, another possibility is to fit asymptotic or sub-asymptotic extreme-value models to high threshold exceedances \citep[see, e.g.,][]{Wadsworth.Tawn:2012b,Huser.Davison:2014,Engelke.etal:2015,Wadsworth.Tawn:2014,Huser.etal:2017,deFondeville.Davison:2018,Huser.Wadsworth:2019}. However, while larger dimensions can generally be tackled based on these models, inference is usually performed using censored likelihood techniques, which involve computationally expensive multi-fold integrals. This prevents their use in really large spatio-temporal problems. To bypass this problem, \citet{deFondeville.Davison:2018} proposed an inference approach based on scoring rules, while \citet{CastroCamilo.Huser:2019} recently proposed an efficient local likelihood estimation approach. Alternatively, more traditional approaches based on (potentially mixtures of) Gaussian-based processes may be exploited to avoid overly prohibitive inference, while accounting for non-stationarity in a flexible way \citep[see, e.g.,][]{Morris.etal:2017,Hazra.etal:2019}. Similarly, machine learning approaches may benefit from high flexibility and fast inference by exploiting parallel computing; see, e.g., \citet{Yu.etal:2017} and the references therein. However, the lack of deep understanding and strong theoretical foundations for these somewhat ad-hoc methods might negatively affect their ability to predict very extreme events.

To foster new research into this direction, a data competition was organized for the EVA 2019 conference, with an application motivated by environmental and ecological considerations. Global warming is affecting the Earth climate year by year, the biggest difference being observable in increasing temperatures in the World Ocean. In particular, coral reefs are increasingly threatened worldwide as they are sensitive to modest increases in background seawater temperature \citep{Cantin.etal:2010}. Studies have shown that persistent high sea temperatures can result in substantial coral bleaching and some cases coral mortality; see, e.g., \citet{McClanahan.etal:2007}. The goal of the EVA 2019 data competition was to analyze and predict the joint tail behavior of extreme sea surface temperature (SST) anomalies for the entire Red Sea, a warm semi-enclosed sea which hosts one of the largest reef systems in the world \citep{Chaidez.etal:2017}, based on a large and high-dimensional dataset. 

The rest of this paper is organized as follows. More details on the data are provided in Section \ref{sec:data}. The precise goal of the data competition, as well as the evaluation criterion and the benchmark prediction, are described in Section~\ref{sec:goal}. The results and the final ranking of the teams are reported in Section~\ref{sec:ranking}, followed by some concluding discussion in Section~\ref{sec:discussion}.

\section{Data}\label{sec:data}
\subsection{Raw data and preprocessing}\label{sec:preproc}
Daily gridded data at a spatial resolution of $1/20^\circ$ (i.e., at an internodal distance of approximately $5.5$km) were produced for the period 1985--2015 by the Operational Sea Surface Temperature and Sea Ice Analysis (OSTIA) system. This data product is based on satellite measurements provided by international agencies, as well as in situ data from ships and buoys, in order to produce accurate SST estimates. The data were provided by GHRSST, Met Office and CMEMS; see \citet{Donlon.etal:2012} for more details.

\begin{figure}[t!]
\centering
\includegraphics[width=\linewidth]{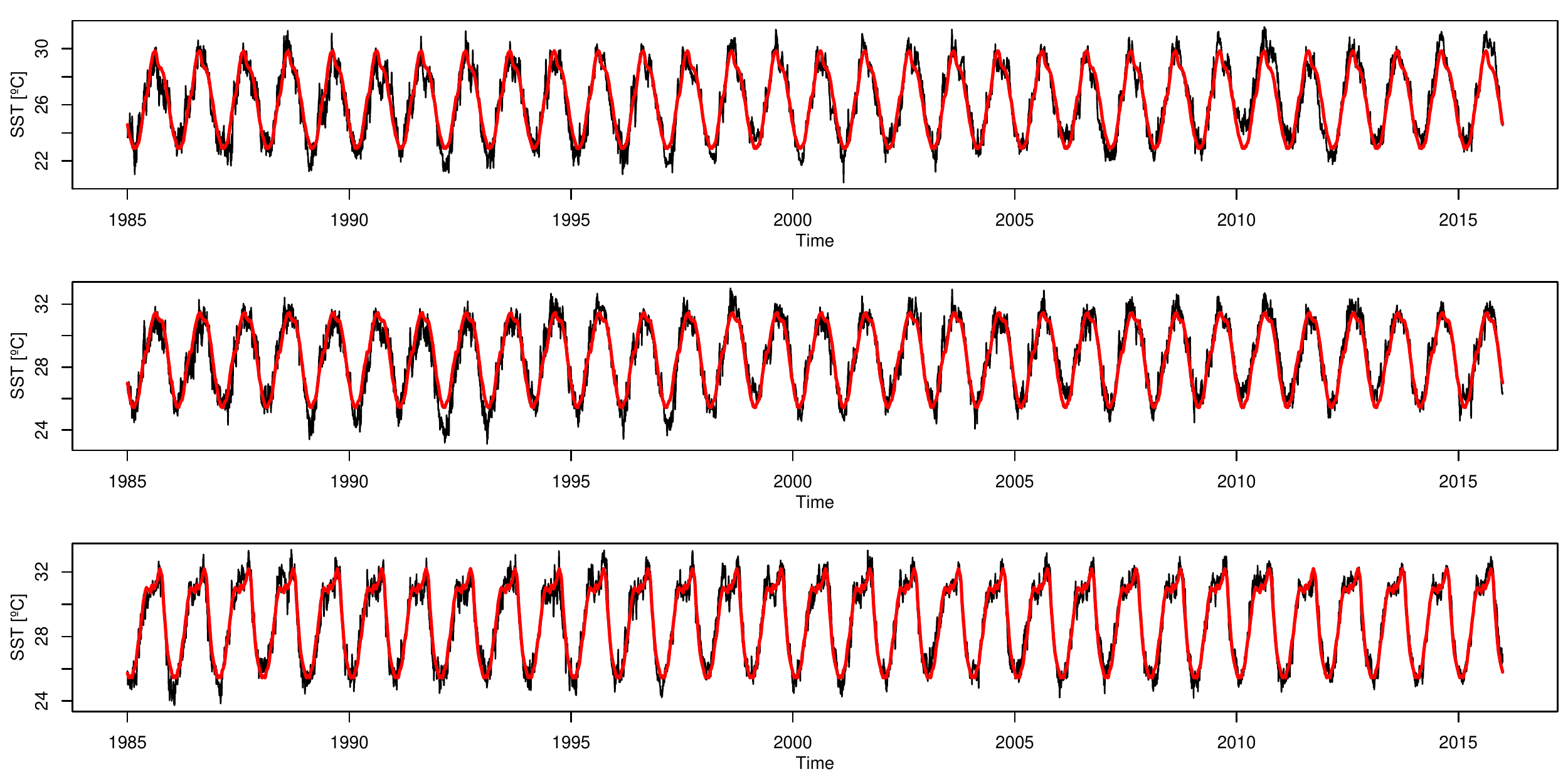}
\caption{SST time series during the entire period 1985--2015 (black) for three locations on the Red Sea (top to bottom panels correspond to North to South locations); see Figure~\ref{fig:SSTmap} for the exact locations. The estimated temperature mean (details in \S\ref{sec:preproc}) is overlaid in red.}\label{fig:SSTtimeseries}
\end{figure}
\begin{figure}[t!]
\centering
\includegraphics[width=\linewidth]{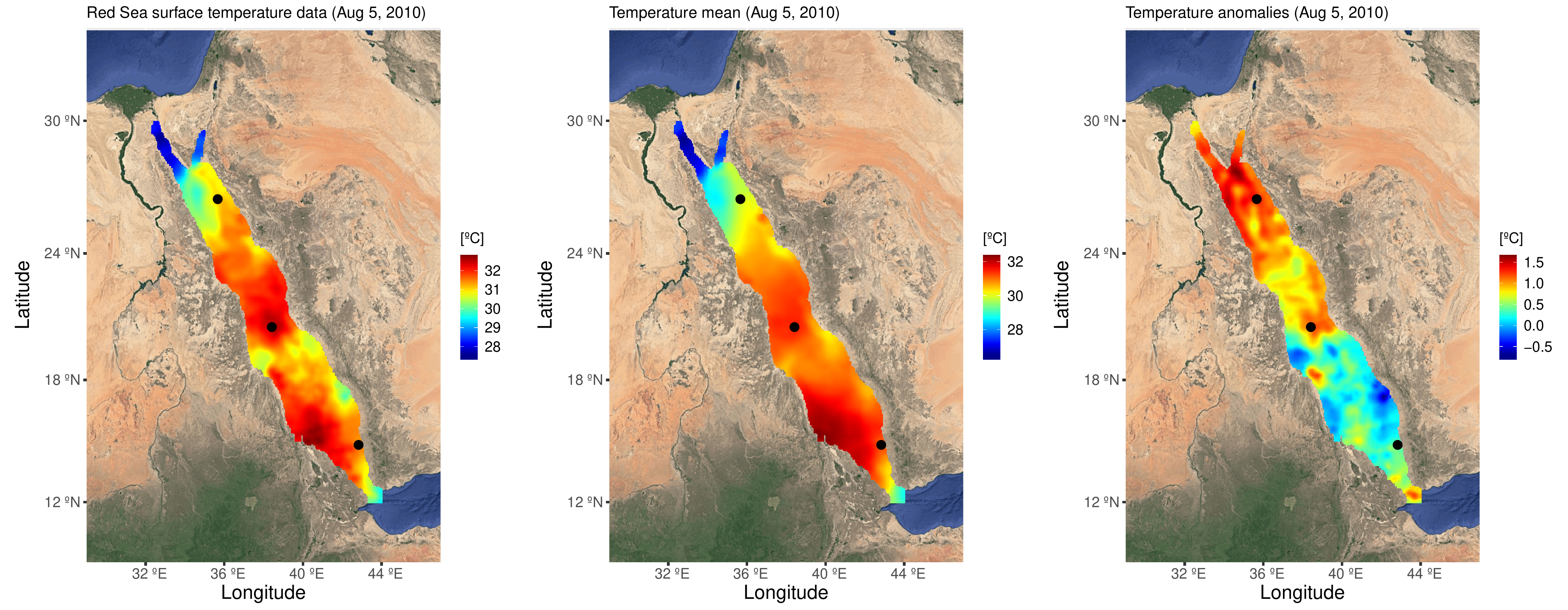}
\caption{SST data $Y(s,t)$ for the entire Red Sea (left), its estimated mean $\hat\mu(s,t)$ (middle) (details in \S\ref{sec:preproc}), and the resulting estimated anomaly $\hat A(s,t)$ (right) for August 5, 2000. Complete time series at three highlighted locations (black dots) are shown in Figure~\ref{fig:SSTtimeseries}.}\label{fig:SSTmap}
\end{figure}

Figure~\ref{fig:SSTtimeseries} shows temperature time series for three locations, while Figure~\ref{fig:SSTmap} displays the spatial variability of the data for August 5, 2010. 
As expected, the data show a clear seasonal pattern and a North--South temperature gradient. There are many ways to deal with this non-stationary behavior. Let $Y(s,t)$ denote the SST process for the Red Sea observed at location $s\in\calS\subset\Real^2$ and time $t\in\calT=\{1,\ldots,T\}$. The Red Sea $\calS$ is discretized into $S=16703$ grid cells, and there are $T=11315$ days in total, giving about 188 million spatio-temporal data points. Here, we decomposed the observed process $Y(s,t)$ into a mean effect $\mu(s,t)$ and the anomaly (or residual component) $A(s,t)$, i.e.,
$$Y(s,t)=\mu(s,t)+A(s,t).$$
To account for spatial variability and seasonality in the mean structure, we simply estimated $\mu(s,t)$ by computing the temperature average for each specific grid cell and each day of the year (by pooling the 31 years together). We then smoothed the estimated mean by computing, for each grid cell separately, a moving average over windows of size one week. This yielded the estimated mean effect $\hat\mu(s,t)$, and the estimated anomalies $\hat A(s,t)$ were finally obtained as
$$\hat A(s,t)=Y(s,t)-\hat\mu(s,t).$$
The teams who participated to the EVA 2019 data competition worked directly with a subset of the estimated anomalies $\hat A(s,t)$, but did not have access to the original data $Y(s,t)$. In this way, the strong non-stationarity in the marginal behavior was (at least partially) removed, in order for the teams to focus more on modeling the spatio-temporal residual process. Figure~\ref{fig:SSTmap} illustrates the temperature process $Y(s,t)$, the estimated mean $\hat\mu(s,t)$, and the estimated anomaly values $\hat A(s,t)$ for August 5, 2010. The estimated mean $\hat\mu(s,t)$ is also plotted as a function of time in Figure~\ref{fig:SSTtimeseries} for three selected spatial locations. As expected, the mean behavior appears to be very smooth, but non-stationary, over both space and time. On the other hand, the anomaly process displays interesting spatial patterns characterized by more rapid fluctuations and local spatial variations.

\begin{figure}[t!]
\centering
\includegraphics[width=0.605\linewidth]{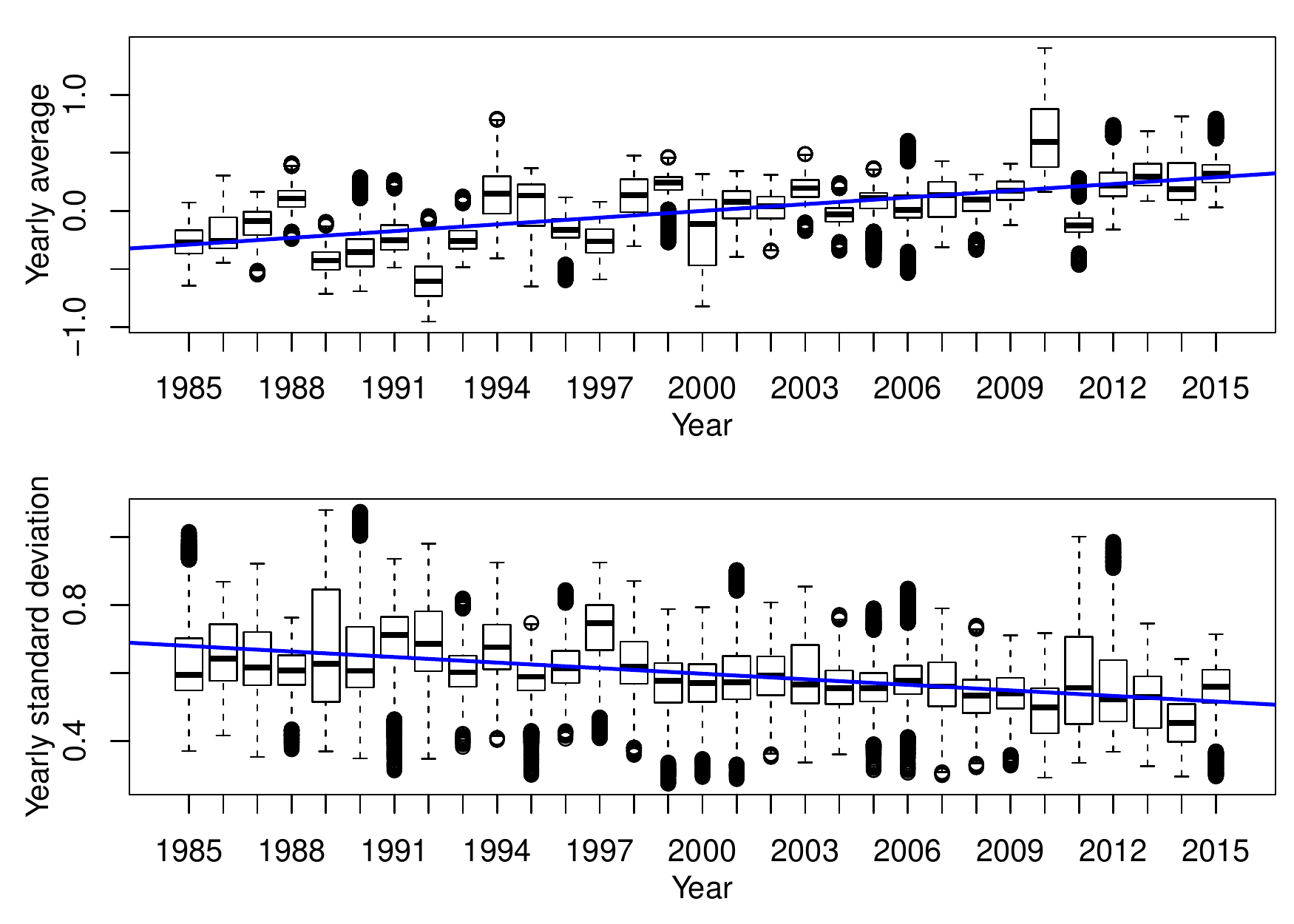}
\includegraphics[width=0.385\linewidth]{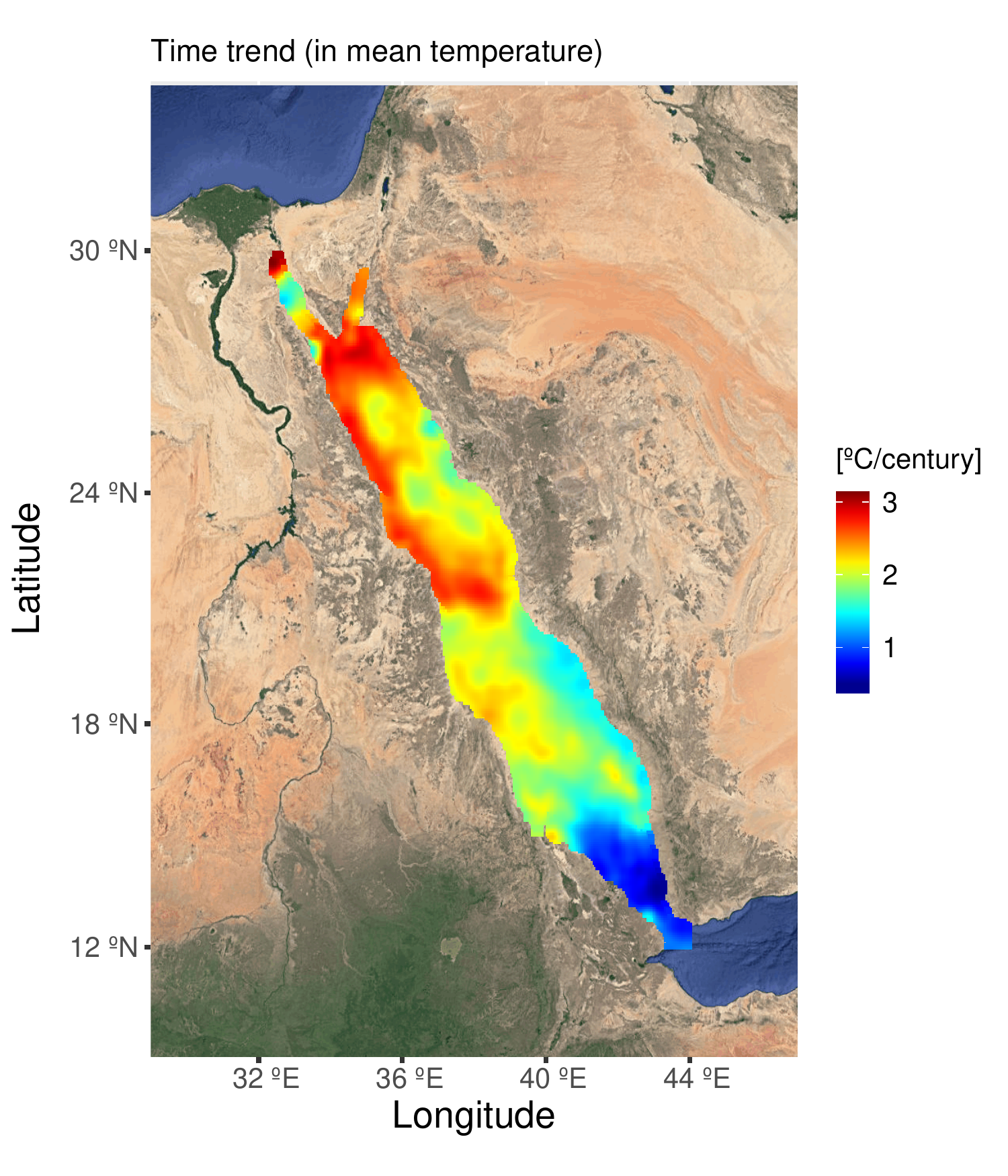}
\caption{Boxplots of yearly anomaly averages (top left) and standard deviations (bottom left) for all spatial locations with the overall linear time trends plotted in blue, and map of spatially-varying trend coefficients [$^\circ$C/century] (for the mean SST), estimated by site-wise linear regressions fitted to the anomaly process (right).}\label{fig:SSTtimetrend}
\end{figure}
While we assumed that the data were roughly year-by-year stationary to estimate the mean $\mu(s,t)$, the yearly boxplots displayed on the left panels of Figure~\ref{fig:SSTtimetrend} reveal however that there are still some small, but clearly visible yearly variations that remain in the mean and standard deviation of the anomaly process $\hat A(s,t)$. Interestingly, while the mean SST increases with time, the standard deviation appears to decrease slightly. Specifically, the mean SST exhibits a statistically significant increase of about $2^\circ$C per century on average, while the standard deviation decreases by about $0.5^\circ$C per century. Moreover, the map on the right panel of Figure~\ref{fig:SSTtimetrend} shows that the time trend varies strongly over space, with more intense warming in the Northern part of the Red Sea. We decided not to remove this time trend, and to let the teams choose whether and how to handle it.

\subsection{Training and validation datasets}

\begin{figure}[t!]
\centering
\includegraphics[width=\linewidth]{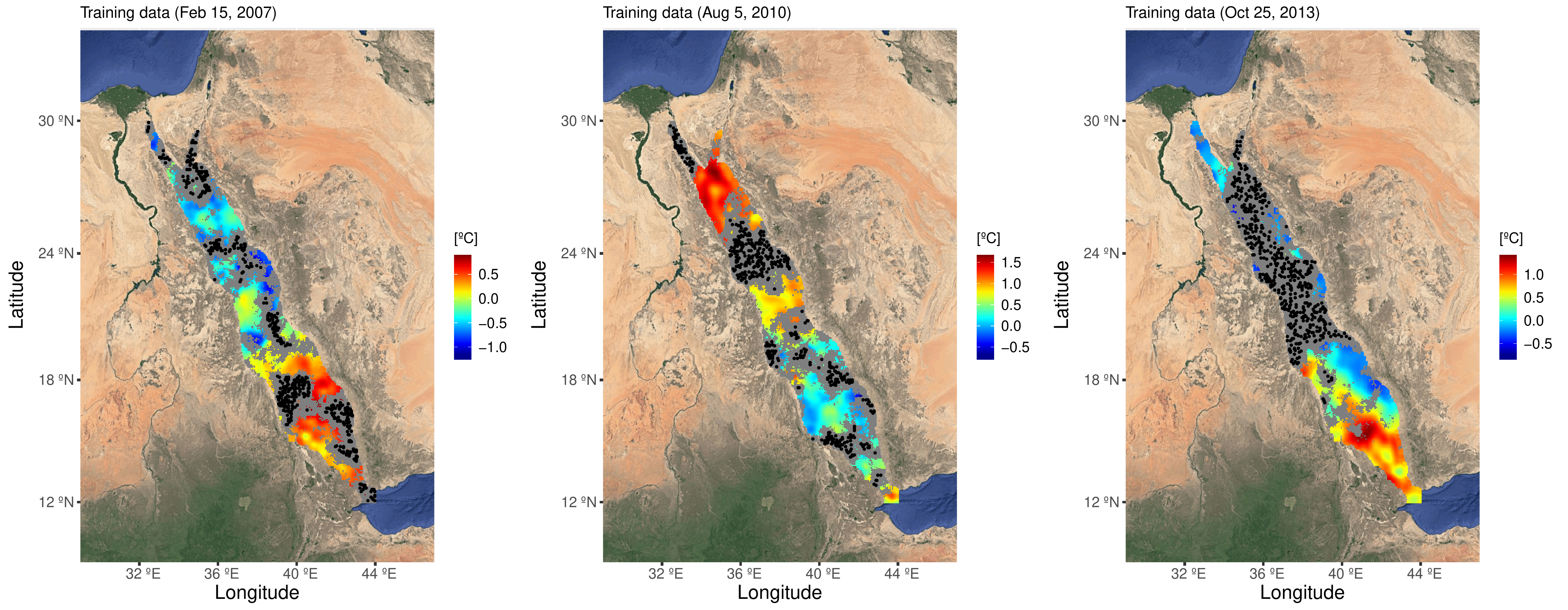}
\caption{Training data, consisting of SST anomalies, here shown for three specific days: February 15, 2007 (left), August 5, 2010 (middle), and October 25, 2013 (right). Grey areas correspond to missing values, while the (overlaid) black dots represent validation points.}\label{fig:SSTmap2}
\end{figure}

For the EVA 2019 data competition, part of the original anomaly data were masked artificially at various places in space and time by introducing missing values (i.e., \tt{NA}s in the statistical software \tt{R}). The missing data mechanism was independent of the observable variables. For each month $\calT_j$, $j=1,\ldots,31\times12=372$ (such that $\cup_{j=1}^{372} \calT_j=\calT$ and $\calT_{j_1}\cap\calT_{j_2}=\emptyset$ for $j_1\neq j_2$), a stationary and isotropic Gaussian random field $Z_j(s)$ was generated over the Red Sea with some chosen spatial correlation structure, and it was then truncated at a suitable level $z_j$, fixed so that the resulting exceedance set $\calM_j={\{s:Z_j(s)>z_j\}}$ identifying missing values for the $j$th month contains a predefined percentage of missing values $\alpha_j$, i.e., $|\calM_j|/S=\alpha_j$. We set $\alpha_j=20\%$ for all months $j$ in the period $1985$--$2006$, and $\alpha_j=60\%$ for the period $2007$--$2015$. Anomaly values $\hat A(s,t)$ were thus treated as missing at all points $(s,t)\in \calM=\cup_{j=1}^{372}(\calM_j\times \calT_j)\subset\calX=\calS\times\calT$. Therefore, the missing value pattern changes every month, and as illustrated in Figure~\ref{fig:SSTmap2}, the data were missing over fairly large spatial areas. Overall, the percentage of missing values was $31.6\%$.

The \emph{training dataset}, made available to the teams, consisted of all non-missing temperature anomaly values, i.e., $\{\hat A(s,t): (s,t)\in \calX_T= \calX\setminus\calM\}$. On the other hand, the \emph{validation dataset} (not available to the teams) consisted of a subset of the missing values indexed by $\calX_V\subset \calM$, comprising $500$ randomly selected spatial locations for the $5$th, $15$th and $25$th of each month during the period $2007$--$2015$. Hence, there were a total of $|\calX_V|=500\mbox{ (locations)}\times 3\mbox{ (days per month)} \times 12 \mbox{ (months)}\times 9\mbox{ (years)} = 162000$ validation points. Clearly, the intersection between the training and validation sets was empty, i.e., $\calX_T\cap\calX_V=\emptyset$. Figure~\ref{fig:SSTmap2} illustrates the training and validation datasets for three chosen days.


\section{Goal, evaluation criterion and benchmark}\label{sec:goal}
\subsection{Main goal of the EVA 2019 data competition}
\begin{figure}[t!]
\centering
\includegraphics[width=\linewidth]{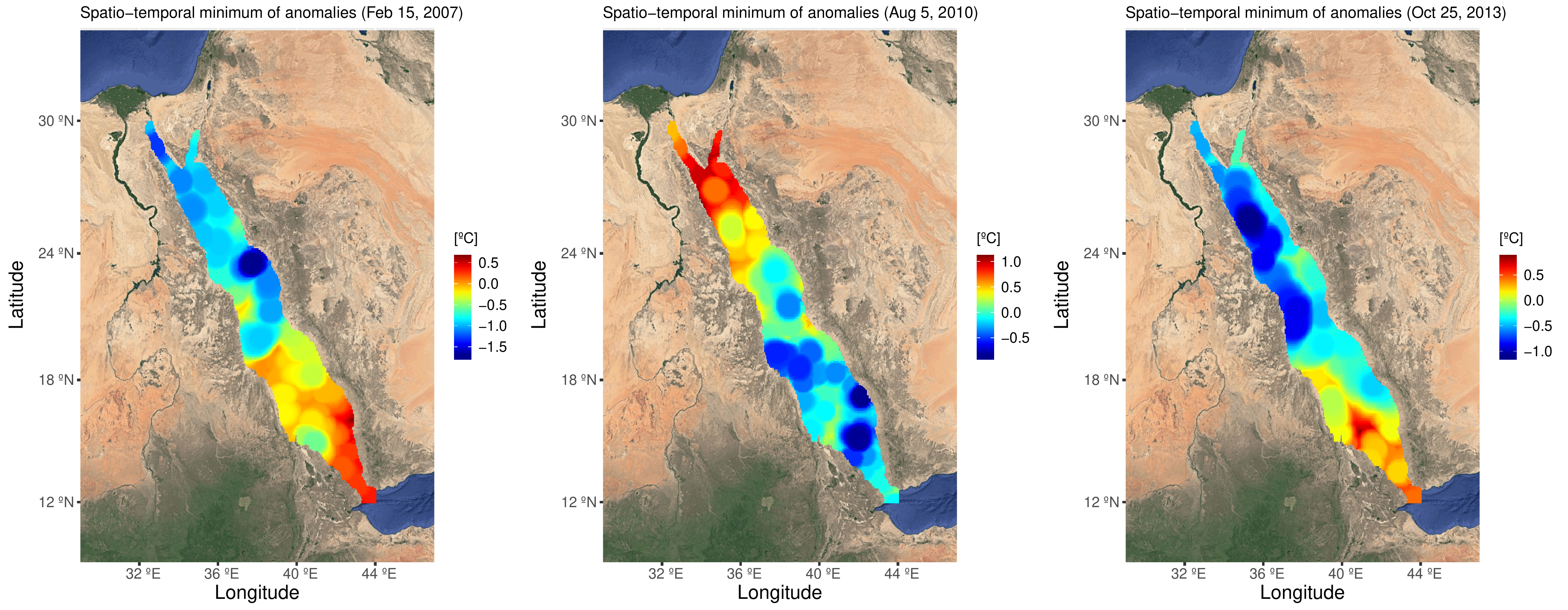}
\caption{Spatio-temporal minimum anomaly process $X(s,t)=\min_{(\tilde{s},\tilde{t})\in\calN(s,t)} \hat A(\tilde{s},\tilde{t})$ in \eqref{eq:Xst} for three specific days: February 15, 2007 (left), August 5, 2010 (middle), and October 25, 2013 (right).}\label{fig:SSTmap3}
\end{figure}

Devastating ecological and environmental degradations are often caused by large-scale extreme temperature events, which are persistently hotter than their usual level and can simultaneously affect an entire region over a prolonged period of time. The main goal of the EVA 2019 data competition, as well as the evaluation criterion (described in Section~\ref{sec:eval}), were designed to reflect this. Specifically, let $\calN(s,t)\subset\calX$ denote a local neighborhood of the spatio-temporal point $(s,t)\in\calX=\calS\times\calT$. Here we considered $\calN(s,t)$ to be a `vertical space-time cylinder' defined as
$$\calN(s,t)=\{\calB(s,r)\times \{t-3,t-2,t-1,t,t+1,t+2,t+3\}\}\cap\calX,$$
where the spatial region $\calB(s,r)$ is a ball centered at location $s$ of radius $r=50$km. 
We then defined spatio-temporal events as extreme at the location $s$ and time $t$ if
\begin{equation}
\label{eq:Xst}
X(s,t)=\min_{(\tilde{s},\tilde{t})\in\calN(s,t)} \hat A(\tilde{s},\tilde{t})>u,
\end{equation}
for some large threshold $u$, where $\hat A(s,t)$ denotes the estimated temperature anomalies. In other words, the definition \eqref{eq:Xst} means that an event is extreme if the SST is simultaneously larger than its mean by $u^\circ{\rm C}$ for at least one week over a (circular) area of radius $50$km. The spatio-temporal minimum anomaly process $X(s,t)$ is displayed in Figure~\ref{fig:SSTmap3} for the same three days as in Figure~\ref{fig:SSTmap2}.
The general objective of the EVA 2019 data competition was to accurately predict the distribution $F_{s,t}$ of $X(s,t)$ defined in \eqref{eq:Xst} for all space-time validation points, i.e., for all $(s,t)\in\calX_V$, paying particular attention to the upper tail.

\subsection{Evaluation criterion}\label{sec:eval}
\begin{figure}[t!]
\centering
\includegraphics[width=0.7\linewidth]{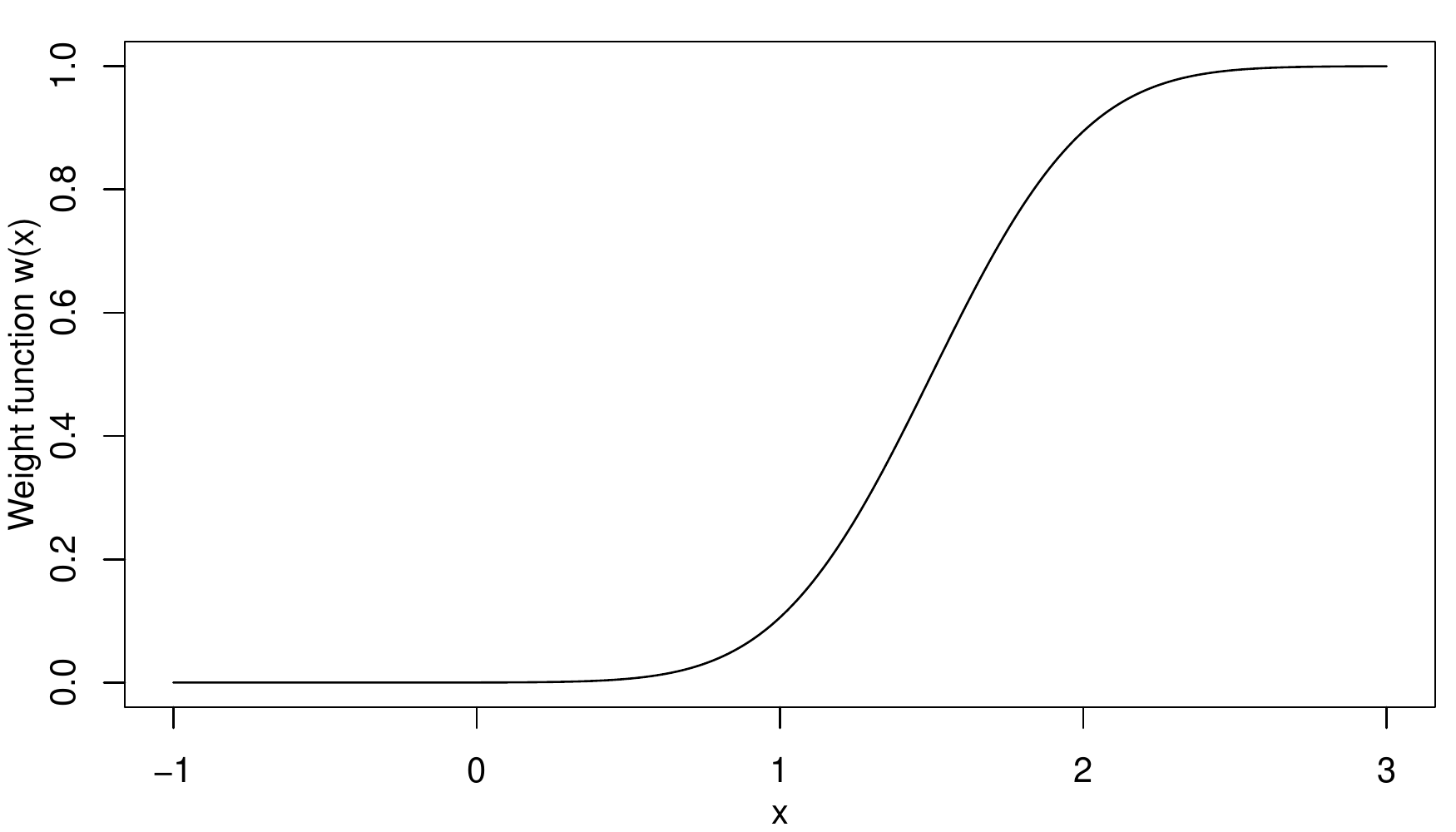}
\caption{Weight function $w(x)=\Phi\{(x-1.5)/0.4\}$ used in \eqref{eq:twCRPS} to compute twCRPS.}\label{fig:weight}
\end{figure}

Let $\hat F_{s,t}$ denote the predicted distribution of $X(s,t)$. 
In order to verify the calibration and sharpness of $\hat F_{s,t}$, while focusing on the upper tail, we used the threshold-weighted continuous ranked probability score (twCRPS) defined as
\begin{equation}
\label{eq:twCRPS}
{\rm twCRPS}(\hat F_{s,t},x_{s,t})=\int_{-\infty}^{\infty}\{\hat F_{s,t}(x)-\mathbb I(x_{s,t}\leq x)\}^2 w(x) {\rm d}x,
\end{equation}
where $\mathbb I(\cdot)$ is the indicator function, $x_{s,t}$ is the observed (realized) value of $X(s,t)$, $w(x)=\Phi\{(x-1.5)/0.4\}$ and $\Phi(\cdot)$ denotes the standard normal distribution. The chosen weight function $w(x)$ is depicted in Figure~\ref{fig:weight}. The twCRPS measure is a proper scoring rule with our choice of weight function $w(x)$; see \citet{Gneiting.Raftery:2007}, \citet{Gneiting.Ranjan:2011}, \citet{Lerch.Thorarinsdottir:2013}, and \citet{Lerch.etal:2017} for more details.

\begin{figure}[t!]
\centering
\includegraphics[width=\linewidth]{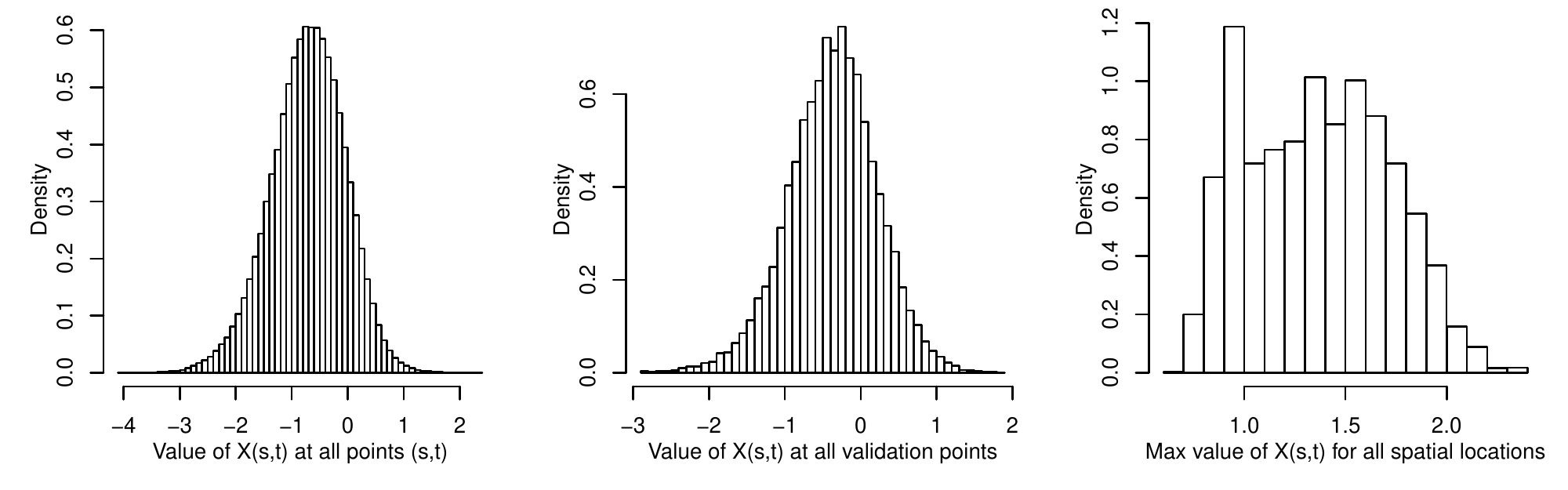}
\caption{Histograms of the true values of the spatio-temporal process $X(s,t)=\min_{(\tilde{s},\tilde{t})\in\calN(s,t)} \hat A(\tilde{s},\tilde{t})$ in \eqref{eq:Xst} for all spatio-temporal locations $(s,t)\in\calX$ (left) and all validation points $(s,t)\in\calX_V$ (middle). The right panel displays the histogram of the maximum values of $X(s,t)$ over time $\calT$, for each spatial location $s\in\calS$.}\label{fig:Histogram}
\end{figure}

Although computing twCRPS requires the full distribution $\hat F_{s,t}$, it puts the emphasis on temperature anomaly values greater than $u\approx1^\circ{\rm C}$. The histograms of true values of $X(s,t)$ shown in Figure~\ref{fig:Histogram} confirm that the weight function $w(x)$ indeed focuses on (very) extreme events.
To compute twCRPS in practice, we restricted the integral in \eqref{eq:twCRPS} to the interval $[-1,3]$ and made the following approximation:
$${\rm twCRPS}(\hat F_{s,t},x_{s,t})\approx\widehat{\rm twCRPS}(\hat F_{s,t},x_{s,t})={1\over 100}\sum_{k=1}^{400}\{\hat F_{s,t}(x^k)-\mathbb I(x_{s,t}\leq x^k)\}^2 w(x^k),$$
where the `design points' were set to $x^k=-1+k/100$, $k=1,\ldots,400$.

The overall prediction accuracy was then assessed by averaging the $\widehat{\rm twCRPS}$ values over the validation set $\calX_V\subset\calX$, i.e., 
$$\bar{\rm twCRPS}={1\over |\calX_V|}\sum_{(s,t)\in\calX_V}\widehat{\rm twCRPS}(\hat F_{s,t},x_{s,t}).$$
The final ranking of the different teams was obtained based on $\bar{\rm twCRPS}$. Lower values of $\bar{\rm twCRPS}$ implied better overall predictions.

\subsection{Benchmark prediction}\label{sec:benchmark}
To calibrate the performance of the different teams and to compare their overall predictive skills with respect to a reference model, we constructed a very simple benchmark by following two basic steps. First, from the training dataset of anomalies $\hat A(s,t)$, we computed the spatio-temporal minimum $x_{s,t}=\min_{(\tilde{s},\tilde{t})\in\calN(s,t)} \hat A(\tilde{s},\tilde{t})$ as in \eqref{eq:Xst} for all points $(s,t)$ that had complete neighborhoods $\calN(s,t)$ (i.e., without any missing values). There were about 40 million space-time locations with complete neighborhoods in total (i.e., about 21\% of the original dataset). Then, assuming stationarity over both space and time, the benchmark prediction $\hat F_{s,t}^{\rm ben}$ was defined for each validation point $(s,t)\in\calX_V$ as the empirical distribution function obtained by pooling all available spatio-temporal minima $x_{s,t}$ together. 

This non-parametric benchmark model has the benefits of being simple to understand, fast to compute, and to rely on minimal assumptions. Moreover, if the stationarity assumption truly holds over space and time, then the benchmark prediction is unbiased and benefits from being estimated from a very large sample (40 million values). However, this assumption of stationarity may be dubious on regions as large as the Red Sea, which suggests that better predictions might be obtained by models capturing spatio-temporal non-stationarity. Moreover, as the benchmark is based on the empirical distribution function, it is expected to perform poorly at estimating very high quantiles, such as those considered in this data competition, despite the large sample size. This was eventually confirmed, since several teams clearly outperformed the benchmark prediction; see the final results in Section~\ref{sec:ranking}.

\section{Final results}\label{sec:ranking}

\begin{table}[t!]
\centering\small
\caption{List of teams who submitted preliminary or final predictions, along with the team members and their affiliations (at the time of the data competition). The teams are listed in alphabetical order.}\label{tab:teams}
\vspace{10pt}
\begin{tabular}{|l|l|l|}\hline
\multicolumn{1}{|c|}{Team name} & \multicolumn{1}{c|}{Team members}  & \multicolumn{1}{c|}{Affiliation} \\ \hline
BeatTheHeat & Daniela Castro-Camilo & KAUST, SA \& \\
& & Univ.\ of Glasgow, UK \\ 
 & Linda Mhalla & HEC Montreal, CA \\ 
 & Thomas Opitz & INRA, Avignon, FR \\ \hline
BlackBox & Domagoj Vlah & Univ.\ of Zagreb, HR \\
 & Tomislav Ivek & Institute of Physics, Zagreb, HR \\ \hline
FNDV & Rapha\"el de Fondeville & EPFL, Lausanne, CH \\ \hline
Jizhi & Jasper Velthoen & TU Delft, Netherlands \\ 
 & Phyllis Wan & Erasmus Univ.\ Rotterdam, NL \\ \hline
Lancaster & Christian Rohrbeck & Lancaster Univ., UK \\ 
 & Emma Simpson & Lancaster Univ., UK \\ 
 & Ross Towe & Lancaster Univ., UK \\ \hline
LancasterTeam2 & Jordan Flett & Lancaster Univ., UK \\ 
 & Robert Shooter & Lancaster Univ., UK \\ 
 & Zak Varty & Lancaster Univ., UK \\
 & Paul Sharkey & JBA Consulting, Skipton, UK \\ \hline
LC2019 & Dan Cheng & TU Delft, NL \\ 
 & Zishun Liu & TU Delft, NL \\ \hline
Multiscale & Seoncheol Park & Seoul National Univ., KR \& \\
 &  & Pacific Climate Impacts Consortium\\
 &  & (PCIC), CA \\
 & Junhyeon Kwon  & Seoul National Univ., KR \\ 
 & Joonpyo Kim & Seoul National Univ., KR \\ 
 & Yaeji Lim & Chung-Ang Univ., Seoul, KR \\ 
 & Hee-Seok Oh & Seoul National Univ., KR \\ \hline
QWER & Gloria Buritica & Sorbonne Univ.\ -- Paris VI, FR \\ \hline
RainbowWarriors & Alexis Hannart & Ouranos, Montreal, CA \\ 
 & Fabien Baeriswyl & McGill Univ., Montreal, CA \\
 & Johanna Neslehova & McGill Univ., Montreal, CA \\ \hline
RedSeaSharksEPFL & Adrian M. C. Hamelink & EPFL, Lausanne, CH \\
-PAVA & Antoine Bourret & EPFL, Lausanne, CH \\ 
& Pierre Vuillecard & EPFL, Lausanne, CH \\
& Victoria Desmarquest & EPFL, Lausanne, CH \\  \hline
RedSeaSharksEPFL & Alejandro De Pascual & EPFL, Lausanne, CH \\
-VASP & Paul Castelain & EPFL, Lausanne, CH\\ 
 & Sami Sellami & EPFL, Lausanne, CH\\
 & Valon Haxhimeri & EPFL, Lausanne, CH\\ \hline
RIKE & Igor Rodionov & Trapeznikov Institute of Control \\ 
 &  & Sciences of Russian Academy of \\
 &  & Sciences, Moscow, RU \\ 
 & Elena Kantonistova & No affiliation \\ \hline
Vitrola & Ruirui Zhang & Univ.\ of Michigan, Ann Arbor, US \\ 
 & Zheng Gao & Univ.\ of Michigan, Ann Arbor, US \\ \hline
\end{tabular}
\end{table}

In total, 22 teams registered for the competition. There were two optional rounds of preliminary predictions, before the teams had to submit their final predictions. Table~\ref{tab:teams} lists all the teams who submitted predictions for the preliminary and/or final rounds. It also includes two teams (BeatTheHeat and RainbowWarriors), who were not able to submit predictions by the final deadline, but who continued to work on it and submitted predictions a few weeks or months later. The competition was fierce, with surprises and turnarounds until the end.

\begin{table}[t!]
\centering\small
\caption{Ranking of all teams for the preliminary rounds 1 (top left) and 2 (top right), the final ranking (bottom left), and the final ranking with the two additional late teams (BeatTheHeat and RainbowWarriors), who submitted their predictions after the deadline (bottom right). The corresponding scores, $10^4\times\bar{\rm twCRPS}$, are also reported. In the first preliminary ranking, the value ``${\rm Score}=+\infty$'' of the 8th team indicates that the submitted predictions were invalid. The winning team is LC2019 (bold font in the final ranking), from the TU Delft, Netherlands.}\label{tab:ranking}
\vspace{10pt}
\begin{tabular}{|c|c|c|p{0.1cm}|c|c|c|}
\multicolumn{3}{c}{Preliminary Ranking 1} & \multicolumn{1}{c}{} & \multicolumn{3}{c}{Preliminary Ranking 2} \\ [5pt]
\multicolumn{7}{c}{}\\[-10pt] \cline{1-3}\cline{5-7}
Rank & Team name & Score &  & Rank & Team name & Score \\ \cline{1-3}\cline{5-7} 
1 & FNDV & 5.068 & & 1 & FNDV & 5.086 \\
2 & Jizhi & 6.815 & & 2 & Lancaster & 5.166 \\
3 & Benchmark & 7.885 & & 3 & LC2019 & 5.324\\
4 & BlackBox & 7.888 & & 4 & Vitrola & 6.601 \\
5 & LC2019 & 12.005 & & 5 & Multiscale & 6.875 \\
6 & LancasterTeam2 & 34.899 & & 6 & RIKE & 7.693 \\
7 & Lancaster & 58.068 & & 7 & BlackBox & 7.859 \\
8 & Multiscale & $+\infty$ & & 8 & Benchmark & 7.885 \\ \cline{1-3}
\multicolumn{3}{c}{} & & 9 & LancasterTeam2 & 7.887 \\
\multicolumn{3}{c}{} & & 10 & RedSeaSharksEPFL & 7.969 \\ 
\multicolumn{3}{c}{} & & & -PAVA/VASP & \\ \cline{5-7}
\multicolumn{7}{c}{} \\ 
\multicolumn{3}{c}{Final Ranking} & \multicolumn{1}{c}{} & \multicolumn{3}{c}{Ranking with the 2 additional late teams} \\ [5pt]
\multicolumn{7}{c}{}\\[-10pt] \cline{1-3}\cline{5-7}
Rank & Team name & Score &  & Rank & Team name & Score \\ \cline{1-3}\cline{5-7}
\textbf{1} & \textbf{LC2019} & \textbf{3.674} & & 1 & BeatTheHeat & 3.279 \\
2 & BlackBox & 4.667 & & 2 & LC2019 & 3.674 \\
3 & RedSeaSharksEPFL & 4.696 & & 3 & RainbowWarriors & 4.471 \\
 & -PAVA &  & & 4 & BlackBox & 4.667 \\
4 & RedSeaSharksEPFL & 4.868 & & 5 & RedSeaSharksEPFL & 4.696 \\
 & -VASP & & & & -PAVA &  \\
5 & FNDV & 5.068 & & 6 & RedSeaSharksEPFL & 4.868 \\
6 & Lancaster & 5.180 & & & -VASP &  \\
7 & Multiscale & 6.799 & & 7 & FNDV & 5.086 \\
8 & Benchmark & 7.885 & & 8 & Lancaster & 5.180 \\
9 & QWER & 7.922 & & 9 & Multiscale & 6.799 \\
10 & Jizhi & 10.545 & & 10 & Benchmark & 7.885 \\ \cline{1-3}
\multicolumn{3}{c}{} & & 11 & QWER & 7.922 \\
\multicolumn{3}{c}{} & & 12 & Jizhi & 10.545 \\ \cline{5-7}
\end{tabular}
\end{table}

Table~\ref{tab:ranking} reports the preliminary and final rankings of the teams, and also the non-official ranking that would have replaced the final ranking, had the two late teams submitted their predictions by the deadline. From the final ranking based on the $\bar{\rm twCRPS}$ measure, the winning team is LC2019 from Delft University of Technology (TU Delft), Netherlands. Congratulations! It is worth noting that the gap with the second team is quite large (20\% reduction in $\bar{\rm twCRPS}$), and that the improvement with respect to the benchmark is substantial (53\% reduction in $\bar{\rm twCRPS}$). Interestingly, the team BeatTheHeat would have been the winners (with an 11\% improvement in $\bar{\rm twCRPS}$ with respect to LC2019), had they submitted their predictions on time. The other late team RainbowWarriors would have been third.

\section{Discussion}\label{sec:discussion}
Because extremes are rare by definition, it is not easy to appropriately compare different methods for modeling and predicting extreme events. The quantile loss function was used in the previous EVA 2017 data competition \citep[see][]{Wintenberger:2018,Opitz.etal:2018}. Here, for the EVA 2019 data competition, we chose instead to rely on the threshold-weighted continuous ranked probability score, which has the advantage of being a proper scoring rule commonly used in probabilistic forecasting, while the weight function can be tailored to put the emphasis on the upper tail. However, there is still an inevitable trade-off in practice: while it may be desirable to assign large weights to very extreme events (potentially beyond the observed data), the scarcity of extremes implies that ``less extreme'' weight functions must be chosen to guarantee a reliable comparison of methods. Our choice of weight function seemed quite reasonable to us, but it remains highly subjective and not ``optimal'' in any sense. Therefore, the results of this competition should be taken with a grain of salt, and we should be careful not to over-interpret the final ranking, as different evaluation criteria might have produced different rankings.

Nevertheless, some general conclusions can still be drawn. Because of the data's high-dimensionality, all teams had to resort to computationally efficient methods. Several teams thus employed machine learning-based approaches or Gaussian-based models, which may not be optimal for capturing heavy tails and strong dependence in spatio-temporal extremes, but yield exceptional speeds-up. The Red Sea SST data considered in this data competition have thin tails, and it would be interesting to compare the same approaches in a heavy-tailed setting where the assumption of multivariate regular variation is likely to hold. More research should be devoted to developing theoretically justified extreme-value methods that scale up to \emph{really} high-dimensional problems. Moreover, the data were highly non-stationary, which also favored flexible models and less traditional approaches. In particular, ``black box'' methods based on machine learning, in which the statistical modeling effort is minimal, but that can be applied in a wide range of settings, performed well overall. Alternatively, local estimation approaches, which can easily handle strong non-stationarity by exploiting the divide-and-conquer strategy, also performed well. 

Recently, \citet{Hazra.Huser:2019} modeled the complete full-resolution SST dataset for the Red Sea by constructing a Bayesian semiparametric mixture model based on low-rank Student $t$ processes. By exploiting empirical orthogonal functions and the resulting low-rank model structure, fully Bayesian inference can be performed at a reasonable computational cost. Moreover, the model captures asymptotic dependence, and its semiparametric specification yields high flexibility to capture strong non-stationarity in space and time. More generally, approaches based on mixtures are natural to use when the observed process is likely to stem from a mixture of processes, e.g., when different days may be driven by different climatic or physical conditions. Such mixture models are useful to automatically cluster days characterized by ``normal conditions'' or ``extreme conditions''. Further research is needed to rigorously handle mixtures of processes in extreme-value modeling, and to precisely assess the predictive skill of the different approaches.

\section*{Acknowledgments}
I would like to thank Bojan Basrak, Hrvoje Planinic and the whole EVA 2019 conference local and scientific committees, for organizing such a successful conference. I also thank Olivier Wintenberger, Alec Stephenson, Holger Rootz\'en and Thomas Mikosch for their support, as well as for helpful discussions and advice on the data competition, and for providing feedback on an early draft of this editorial. Finally, I thank and congratulate all teams, without whose active and positive participation this competition would not have taken place.

\baselineskip=20pt


\baselineskip 10pt

\end{document}